\documentclass[11pt,a4paper]{article}
\pdfoutput=1
\usepackage{a4wide}
\usepackage{amsfonts}
\usepackage{amssymb}
\usepackage{amsmath}
\usepackage{graphicx}
\usepackage{booktabs}
\usepackage{array}
\usepackage{color}
\usepackage{ulem}
\usepackage[small,bf]{caption}
\usepackage[usenames,dvipsnames]{xcolor}
\usepackage{subfigure}
\usepackage{verbatim}
\usepackage{dsfont}
\usepackage{bbm}
\usepackage[backend=bibtex8,sorting=none,natbib=true,maxbibnames=10,style=numeric-comp,block=ragged]{biblatex}
\usepackage{hyperref}
\usepackage{listings}
\bibliography{literature}

\lstdefinestyle{cpp}{
  belowcaptionskip=1\baselineskip,
  breaklines=true,
  xleftmargin=\parindent,
  language=C,
  showstringspaces=false,
  basicstyle=\footnotesize\ttfamily,
  morekeywords={gauge,yukawa,self,Eigen,std,complex,class,cout,endl,using,namespace,boost,public,private,bool,template,typename},
  keywordstyle=\bfseries\color{green!40!black},
  classoffset=0,
  commentstyle=\itshape\color{red!40!black},
  identifierstyle=\color{blue},
  stringstyle=\color{orange},
}

\hypersetup{%
  colorlinks=true, linktocpage=true, pdfstartpage=1, pdfstartview=FitV,
  breaklinks=true, pageanchor=true,
  pdfpagemode=UseNone,
  plainpages=false, bookmarksnumbered, bookmarksopen=true, bookmarksopenlevel=1,
  hypertexnames=true, pdfhighlight=/O,
  urlcolor=RoyalBlue, linkcolor=Maroon, citecolor=RoyalBlue,
  pdftitle={\texttt{RGE++}},
  pdfauthor={Thomas Deppisch and Florian Herren},
  pdfsubject={},
  pdfkeywords={},
  pdfcreator={pdfLaTeX},
  pdfproducer={LaTeX with hyperref}
}

\begin{document}
\lstset{style=cpp}

\begin{titlepage}

\vspace*{-15mm}
\begin{flushright}
TTP19-025\\
P3H-20-085
\end{flushright}
\vspace*{0.7cm}

\begin{center}
{ \bf\LARGE \texttt{RGE++}\\ A \texttt{C++} library to solve renormalisation group equations in quantum field theory}
\\[8mm]
Thomas Deppisch$^{\, a}$ \footnote{E-mail: \texttt{thomas.deppisch@dwd.de}} ,
Florian Herren$^{\, b}$ \footnote{Corresponding Author, E-mail: \texttt{florian.herren@kit.edu}}
\\[1mm]
\end{center}
\vspace*{0.50cm}
\centerline{$^{a}$ \it German Weather Service (DWD)}
\centerline{\it Research and Development --- Data Assimilation}
\centerline{\it Frankfurter Stra\ss{}e 135, D-73067 Offenbach am Main, Germany}
\vspace*{0.2cm}
\centerline{$^{b}$ \it Institut f\"ur Theoretische Teilchenphysik}
\centerline{\it Karlsruhe Institute of Technology (KIT)}
\centerline{\it Wolfgang-Gaede-Stra\ss{}e 1, D-76128 Karlsruhe, Germany}
\vspace*{0.2cm}

\begin{abstract}
In recent years three-, four- and five-loop beta functions have been computed for various phenomenologically interesting models.
However, most of these results have not been implemented in easy to use software packages. \texttt{RGE++} bridges
this gap by providing a flexible, template-based, \texttt{C++} library to solve renormalisation group equations.
Furthermore, we implement the available beta functions for the Standard Model, the minimal supersymmetric
extension of the Standard Model and two-Higgs-doublet models, as well as right-handed neutrino extensions of the former two.
\end{abstract}

\end{titlepage}

\setcounter{footnote}{0}

\section{Introduction}
When renormalised in the modified minimal subtraction ($\overline{\mathrm{MS}}$) or modified dimensional reduction ($\overline{\mathrm{DR}}$) schemes,
parameters and fields in quantum field theories depend on the renormalisation scale $\mu$. The change of a parameter with respect to a change of $\mu$
is encoded in anomalous dimensions, or, in the case of coupling constants, the beta functions
\begin{align}
\frac{\mathrm{d}}{\mathrm{d}\ln\mu}g_i = \beta_g\left(\left\{g_j\right\}\right)~,
\label{eq::beta}
\end{align}
where the $g_i$ denote any dimensionless coupling of the theory under consideration and $\left\{g_j\right\}$ denotes the set of
all couplings of the theory.
Beta functions play a crucial role in connecting parameters at different scales, for example when comparing measurements of the
strong coupling constant at different energy scales at the Z-Boson mass $\mu = M_Z$. Other applications include the study of
Grand Unified Theories (GUTs) by extrapolating the measured couplings at low scales to the unification scale or the generation
of supersymmetric (SUSY) spectra from high-scale input.

Usually, Eq.~(\ref{eq::beta}) is solved numerically and over the years many public programs appeared which implement
anomalous dimensions and beta functions of various models. Most of them implement more features than the scope of \texttt{RGE++}, but are rather specific to a certain model. 
Here we provide a short overview of programs or libraries which are relevant in the context of renormalisation group equations (RGEs):
For pure QCD at the highest perturbative orders available there is \texttt{(C)RunDec} \cite{Chetyrkin:2000yt,Schmidt:2012az,Herren:2017osy}. \texttt{SMDR} \cite{Martin:2019lqd} and \texttt{mr} \cite{Kniehl:2016enc}
implement running and matching relations in the full Standard Model,
while REAP \cite{Antusch:2005gp} implements beta functions in the SM, MSSM and 2HDMs including right-handed neutrinos. For supersymmetric theories there is a plethora of spectrum generators,
which all implement various models, such as \texttt{SOFTSUSY} \cite{Allanach:2001kg}, \texttt{SuSpect} \cite{Djouadi:2002ze}, \texttt{SPheno} \cite{Porod:2003um,Porod:2011nf},
\texttt{Flexible SUSY} \cite{Athron:2014yba,Athron:2017fvs}. Furthermore, there are programs implementing the general two-loop beta functions
\cite{Machacek:1983tz,Machacek:1983fi,Machacek:1984zw,Jack:1982hf,Jack:1982sr,Jack:1984vj} and anomalous dimensions of the dimensionful couplings \cite{Luo:2002ti,Sartore:2020pkk}, such
as \texttt{SARAH} \cite{Staub:2013tta}, \texttt{PyR@TE} \cite{Lyonnet:2016xiz,Sartore:2020gou} and \texttt{ARGES} \cite{Litim:2020jvl}\footnote{\texttt{PyR@TE} and \texttt{ARGES} also implement the results for three-loop gauge coupling beta functions \cite{Poole:2019kcm}}.

The goal of \texttt{RGE++} is much simpler: to provide an easy to use, template-based \texttt{C++} class library, capable of numerically solving rather generic
RGEs, which can be used from within other programs. Furthermore, we want to provide various higher-order results, since many of
the aforementioned programs only implement the relevant beta functions and anomalous dimensions up to two loops. 

Given recent developments
such as the extraction of the gauge coupling beta function in a general quantum field theory at three loops \cite{Poole:2019txl,Poole:2019kcm}, the computation
of the gauge coupling beta functions in the SM at four loops \cite{Davies:2019onf}, as well as the identification of all colour structures appearing in the four-loop
gauge and three-loop Yukawa coupling beta functions, we can expect a plethora of new results for various models. A flexible code such as \texttt{RGE++}
will be able to incorporate such new results quickly.

The remainder of this paper is structured as follows: In Section \ref{sec::manual} we present the base classes and discuss the models implemented.
Section \ref{sec::examples} showcases the usage of \texttt{RGE++} with four examples.
A brief summary, as well as an outlook on further developments can be found in \ref{sec::sum}.

\section{\label{sec::manual}The RGE++ classes and functions}
In this section we discuss the functionality of \texttt{RGE++} and its implementation. Readers who are more interested in actual examples are encouraged to skip ahead to
Section \ref{sec::examples} and come back to this section afterwards.

First, in Sec.~\ref{sec::installation}, we describe the installation and the dependencies of \texttt{RGE++},
followed by a brief overview about the implemented models in Sec.~\ref{sec::overview}.
The description of the base classes all models are derived from can be found in Sec.~\ref{sec::base}. In Sec.~\ref{sec::models} we discuss the implementation of the various models
and fix the conventions. We conclude with documenting auxiliary functions in Sec.~\ref{sec::auxiliary}.

\subsection{\label{sec::installation}Installation and usage}
The code can be obtained from \url{https://github.com/Herren/RGEpp} and is structured as follows: the folders \texttt{include} and \texttt{src} contain the header and source files of the basic classes, respectively.
\texttt{models} contains the code for all implemented models, while \texttt{examples} contains the code for various examples, including those documented in the following.

To use \texttt{RGE++} an installation of the libraries \texttt{Eigen3} \cite{eigenweb} and \texttt{odeint} \cite{Ahnert:2011} is necessary.
\texttt{Eigen3} provides template-based implementations of vector and matrix classes. \texttt{odeint} implements standard algorithms for numerically solving differential equations
and can operate on the matrix classes provided by \texttt{Eigen3}.
They can be obtained by following the instructions in the \texttt{README} file. For building the examples, the variables \texttt{EIGENPATH} and \texttt{ODEINTPATH} in the \texttt{Makefile} might need to be changed accordingly.

\subsection{\label{sec::operation} General structure}
The source code of \texttt{RGE++} is distributed over four folders:
\begin{itemize}
\item \texttt{include}: contains header files of template and auxiliary classes
\item \texttt{src}: contains source files of auxiliary classes
\item \texttt{models}: contains model implementations
\item \texttt{examples}: contains examples showcasing the use of \texttt{RGE++}
\end{itemize}
All models are derived from two template classes with a flexible number of gauge and quartic couplings: \texttt{base} and \texttt{nubase}. The latter extends the former by additional matrices to account for neutrino Yukawa couplings
and Majorana masses. Both template classes provide all routines and operators for solving the RGEs numerically using \texttt{odeint}.

Specific models are derived from either of the two base classes and implement the operator
\begin{lstlisting}
      void operator()(const class & X, class & dX, const double)
\end{lstlisting}
This operator is called by \texttt{odeint} at each step of the numerical integration, writing the derivative of \texttt{X} into \texttt{dX}.
Thus for each model, this operator needs to implement the beta functions of the model. A more detailed discussion is given in Sec.~\ref{sec::examples}.

\subsection{\label{sec::overview}Overview of implemented models}
\texttt{RGE++} implements the SM beta functions for gauge couplings, Yukawa matrices and the scalar quartic coupling up to three loops \cite{Machacek:1983tz,Machacek:1983fi,Machacek:1984zw,
                               Mihaila:2012fm,Mihaila:2012pz,Bednyakov:2012rb,Chetyrkin:2012rz,Bednyakov:2012en,Bednyakov:2014pia,Herren:2017uxn,Chetyrkin:2013wya,Bednyakov:2013eba,Bednyakov:2013cpa}.
Furthermore, the four-loop corrections to the gauge coupling beta functions \cite{Davies:2019onf} are available\footnote{Note that in \cite{Davies:2019onf} only third generation Yukawa couplings are taken into account.}.
The non-supersymmetric extensions implemented are the SM including right-handed neutrinos at two-loop order \cite{Machacek:1983tz,Machacek:1983fi,Machacek:1984zw, Antusch:2005gp, Grzadkowski:1987tf}, as well as
the $\mathbb{Z}_2$-symmetric two-Higgs-doublet models at three-loop order for gauge coupling and Yukawa matrices and two-loop order for the quartic couplings
\cite{Machacek:1983tz,Machacek:1983fi,Machacek:1984zw, Herren:2017uxn, Chowdhury:2015yja}.
The beta functions for the minimal supersymmetric extension of the SM are implemented up to three-loop order \cite{Martin:1993zk,Ferreira:1996ug} and its extension with right-handed (s)neutrinos at two-loop order
\cite{Martin:1993zk,Antusch:2005gp,Grzadkowski:1987wr}. All implemented models employ $\mathrm{SU}(5)$ normalization for the $\mathrm{U}(1)_\mathrm{Y}$ gauge coupling $g_1$.

A full list, including all references, can be found in Table \ref{tbl:models}.

\begin{table}
  \centering
  \begin{tabular}{lll} \toprule
    file name & loop order & publications \\\midrule
    \texttt{sm.h}  & 2 & \cite{Ahnert:2011,eigenweb,Machacek:1983tz,Machacek:1983fi,Machacek:1984zw} \\
    \texttt{sm.h}  & 3 & \cite{Ahnert:2011,eigenweb,Machacek:1983tz,Machacek:1983fi,Machacek:1984zw,
                               Mihaila:2012fm,Mihaila:2012pz,Bednyakov:2012rb,Chetyrkin:2012rz,Bednyakov:2012en,Bednyakov:2014pia,Herren:2017uxn,Chetyrkin:2013wya,Bednyakov:2013eba,Bednyakov:2013cpa} \\
    \texttt{sm.h}  & 4 & \cite{Ahnert:2011,eigenweb,Machacek:1983tz,Machacek:1983fi,Machacek:1984zw, 
                               Mihaila:2012fm,Mihaila:2012pz,Bednyakov:2012rb,Chetyrkin:2012rz,Bednyakov:2012en,Bednyakov:2014pia,Herren:2017uxn,Chetyrkin:2013wya,Bednyakov:2013eba,Bednyakov:2013cpa, Davies:2019onf} \\
    \midrule
    \texttt{mssm.h} & 2 & \cite{Ahnert:2011,eigenweb,Martin:1993zk} \\
    \texttt{mssm.h} & 3 & \cite{Ahnert:2011,eigenweb,Martin:1993zk,Ferreira:1996ug,Harlander:2009mn} \\
    \texttt{nusm.h}  & 2 & \cite{Ahnert:2011,eigenweb,Machacek:1983tz,Machacek:1983fi,Machacek:1984zw, Antusch:2005gp, Grzadkowski:1987tf} \\
    \texttt{numssm.h} & 2 & \cite{Ahnert:2011,eigenweb, Martin:1993zk,Antusch:2005gp,Grzadkowski:1987wr} \\
    \texttt{thdmi.h} & 3 & \cite{Ahnert:2011,eigenweb, Machacek:1983tz,Machacek:1983fi,Machacek:1984zw, Herren:2017uxn, Chowdhury:2015yja}  \\
    \texttt{thdmii.h} & 3 & \cite{Ahnert:2011,eigenweb, Machacek:1983tz,Machacek:1983fi,Machacek:1984zw,Herren:2017uxn, Chowdhury:2015yja}  \\
    \texttt{thdmx.h} & 3 & \cite{Ahnert:2011,eigenweb, Machacek:1983tz,Machacek:1983fi,Machacek:1984zw,Herren:2017uxn, Chowdhury:2015yja}  \\
    \texttt{thdmy.h} & 3 & \cite{Ahnert:2011,eigenweb, Machacek:1983tz,Machacek:1983fi,Machacek:1984zw,Herren:2017uxn, Chowdhury:2015yja}  \\
    \midrule
    \texttt{ckm.h} & any & \cite{Ahnert:2011,eigenweb,Antusch:2005gp}  \\
    \texttt{pmns.h} & any & \cite{Ahnert:2011,eigenweb,Antusch:2005gp}  \\
    \texttt{sm\_example.cpp} & 2 & \cite{Ahnert:2011,eigenweb,Deppisch:2018flu,Antusch:2005gp,Machacek:1983tz,Machacek:1983fi,Machacek:1984zw} \\
    \bottomrule
  \end{tabular}
  \caption{\label{tbl:models}Relevant literature for each implemented model that should be cited together with this work when using \texttt{RGE++} in scientific publications.}
\end{table}

\subsection{\label{sec::base} Base classes}
\paragraph{eigentypes.h}
sets the type definitions
\begin{lstlisting}
  template<int n>
  using gauge = Eigen::Matrix<double, n,1>;
   
  using yukawa = Eigen::Matrix<std::complex<double>, 3,3>;

  template<int n>
  using self = Eigen::Matrix<std::complex<double>, n,1>;
\end{lstlisting}
where \texttt{gauge} and \texttt{self} represent $n$-dimensional vectors of gauge and quartic couplings, while \texttt{yukawa} represents fixed-size quadratic Yukawa matrices.
Furthermore, for each of the aforementioned types, the functions
\begin{lstlisting}
  typename abs(typename);  // returns the coefficient-wise absolute values
  bool isnotnan(typename); // checks whether a coupling is 'NaN'
\end{lstlisting}
are defined, which are used internally in the routines solving the RGE numerically.
The types defined in this header file constitute the basic data types on which the various classes are built.

\paragraph{base.h}
defines the class \texttt{base}, which serves as a base class from which all model classes with three Yukawa matrices are derived. The following model parameters are defined as \texttt{public} members:
\begin{lstlisting}
  gauge<n> g;      // vector of gauge couplings
  self<m> La;      // Higgs self-coupling(s)
  yukawa Yu,Yd,Ye; // Yukawa matrices
\end{lstlisting}
Here \texttt{n} and \texttt{m} are the template parameters of \texttt{base}, specifying the number of gauge and quartic couplings of the model, respectively.
The loop order for the beta functions is via the \texttt{protected} member
\begin{lstlisting}
  unsigned int nloops; // loop order of the beta functions
\end{lstlisting}
In all models derived from \texttt{base} \texttt{nloops} defaults to \texttt{2}.

\texttt{base.h} defines the following member functions of \texttt{base} as public:
\begin{lstlisting}
  void setZero() // sets g, La, Yu, Yd, Ye to zero
  bool check() // 'true' if no Landau poles appear and no member is 'NaN'
  void setNloops(const unsigned int) // sets the value of nloops
  unsigned int getNloops() // returns the current value nloops
\end{lstlisting}
Besides, \texttt{base} provides all vector space operations needed for the numerical integration with \texttt{odeint}. The classes \texttt{sm}, \texttt{thdmi/ii/x/y}, \texttt{mssm} are derived from \texttt{base}.

\paragraph{nubase.h}
defines the class \texttt{nubase} from which all model classes with four Yukawa matrices, the Wilson coefficient of the Weinberg operator, as well as a Majorana mass for right-handed neutrinos are derived.
In analogy to \texttt{base}, it has the following model parameters as \texttt{public} members:
\begin{lstlisting}
  gauge<n> g;         // vector of gauge couplings
  self<m> La;         // Higgs self-coupling(s)
  yukawa Yu,Yd,Ye,Yn; // Yukawa matrices
  yukawa Ka,Mn;       // Wilson coefficient of the Weinberg operator,
                      // neutrino mass matrix   
\end{lstlisting}
The loop order for the beta functions is set via the \texttt{protected} member
\begin{lstlisting}
  unsigned int nloops; // loop order of the beta functions
\end{lstlisting}
In all models derived from \texttt{nubase} \texttt{nloops} defaults to \texttt{2}.
In analogy to \texttt{base.h}, \texttt{nubase.h} defines the following public member functions of \texttt{nubase}
\begin{lstlisting}
  void setZero(); // sets g, La, Yu, Yd, Ye, Yn, Ka, Mn to zero
  bool check(); // 'true' if no Landau poles appear and no member is 'NaN'
  void setNloops(const unsigned int); // sets the value of nloops
  unsigned int getNloops(); // returns the current value nloops
\end{lstlisting}
and provides all vector space operations needed for the numerical integration with \texttt{odeint}.

Furthermore, the class \texttt{nubase} supports the decoupling of heavy right-handed neutrino. The tree-level matching is performed as described in \cite{Antusch:2005gp} according to
\begin{equation}
    \kappa_{ij}^{\rm low} = \kappa_{ij}^{\rm high} + 2\, \bigl(Y_\nu^T\bigr)_{in} M_n^{-1} \bigl(Y_\nu\bigr)_{nj} \;,
\end{equation}
in the basis where \texttt{Mn} is diagonal. Here, $\kappa$ corresponds to \texttt{Ka}, the Wilson coefficient of the Weinberg operator.
To this end, the class \texttt{nubase} implements the function
\begin{lstlisting}
  void integrate_out(int n);
\end{lstlisting}
which integrates out the $n^\mathrm{th}$ generation (counting from zero) of right-handed neutrinos. 
Afterwards, the $n^\mathrm{th}$ row of $Y_\nu$ (counting from zero) is set to zero.
In addition, 
\begin{lstlisting}
  Eigen::Vector3d logthresholds();
\end{lstlisting}
returns the logarithm of the eigenvalues of the right-handed neutrino mass matrix, while
\begin{lstlisting}
  yukawa getML(double vev);
  yukawa getML(double vev, double tanb);
\end{lstlisting}
computes the masses of the light neutrinos.
The classes \texttt{nusm} and \texttt{numssm} are derived from \texttt{nubase}.

\subsection{\label{sec::models} Implemented models}
\paragraph{sm.h}
implements the class \texttt{sm}, which is derived as \texttt{public} from \texttt{base} and therefore has all members and member functions of \texttt{base}. A \texttt{sm} object can be constructed by
\begin{lstlisting}
  sm() : base<3,1>() {};
  
  sm(const gauge<3> g_in, const std::complex<double> La_in,
     const yukawa Yu_in, const yukawa Yd_in, const yukawa Ye_in,
     const int nloops_in)
   : base<3,1>(g_in, Yu_in, Yd_in, Ye_in, nloops_in) { La[0] = La_in; };
   
  sm(const gauge<3> g_in, const std::complex<double> La_in,
     const yukawa Yu_in, const yukawa Yd_in, const yukawa Ye_in)
   : base<3,1>(g_in, Yu_in, Yd_in, Ye_in) { La[0] = La_in; };
   
  sm(const base<3,1> &X) : base<3,1>(X) {};
\end{lstlisting}
The first constructor initializes the values of all couplings to $0$, whereas in the second and third version the couplings are passed as arguments. The third version sets the number of loops to $2$.
Finally, the fourth version serves as copy-constructor.
Furthermore, \texttt{sm} provides the operator
\begin{lstlisting}
  void operator()(const sm &X, sm &dX, const double);
\end{lstlisting}
which is implemented in \texttt{sm.cpp}. This operator contains the actual implementation of the beta functions and is called by the \texttt{odeint} routines when running from one scale to another.

In the implementation of the beta functions, the `right-left-convention' for the Yukawa interactions is used
\begin{equation} \label{eq:sm_yuk}
    \mathcal{L}_{\mathrm{Yukawa}}^{\mathrm{SM}} = - Y_e^{ij}\; \overline{e_{R}}_{i} \, \phi^\dagger \cdot L_j
  - Y_d^{ij}\; \overline{d_{R}}_{i} \, \phi^\dagger \cdot Q_j
  - Y_u^{ij}\; \overline{u_{R}}_{i} \, Q_j^T \cdot \varepsilon \cdot \phi + \mathrm{H.c.}~,
\end{equation}
where $\varepsilon = \left(\begin{smallmatrix} 0 & 1 \\ -1 & 0 \end{smallmatrix}\right)$, $\cdot$ denotes matrix multiplication and the charge conjugated Dirac spinor is $\psi^C = -\mathrm{i}\gamma^2\gamma^0\psi^T$. For the Higgs self-coupling $\lambda$ we have\footnote{Thus, $\lambda = m_h^2/v^2$ with $v=174.14\,\mathrm{GeV}$.}
\begin{equation} \label{eq:sm_la}
  \mathcal{L}_{\mathrm{self}} = -\frac{\lambda}{4} (\phi^\dagger\phi)^2
\end{equation}
\texttt{La} is implemented in \texttt{base} as a vector of \texttt{std::complex<double>}. In \texttt{sm}, therefore, all entries but \texttt{La[0]} are set to zero at construction.

\paragraph{thdm.h}
implements the classes \texttt{thdmi}, \texttt{thdmii}, \texttt{thdmx} and \texttt{thdmy}, which denote the four $\mathbb{Z}_2$-symmetric 2HDMs.
They are derived as \texttt{public} from \texttt{base} and therefore have all members and member functions of \texttt{base}.
A Type-I 2HDM, for example, can be initialized by
\begin{lstlisting}
  thdmi() : base() {};
 
  thdmi(const gauge g_in, const self La_in, const yukawa Yu_in,
        const yukawa Yd_in, const yukawa Ye_in, const int nloops_in)
      : base(g_in, La_in, Yu_in, Yd_in, Ye_in, nloops_in) {};
      
  thdmi(const gauge g_in, const self La_in, const yukawa Yu_in,
        const yukawa Yd_in, const yukawa Ye_in)
      : base(g_in, La_in, Yu_in, Yd_in, Ye_in) {};
      
  thdmi(const base &X) : base(X) {};
\end{lstlisting}
The only difference w.r.t. the SM case is that the self-couplings now form a vector.

Like in the SM case, we use the `right-left-convention' and take $\phi_2$ to be the SM-like doublet. So in the Type-I 2HDM
\begin{equation} \label{eq:thdm_yuk}
    \mathcal{L}_{\mathrm{Yukawa}}^{\mathrm{2HDM,I}} = - Y_e^{ij}\; \overline{e_{R}}_{i} \, \phi_2^\dagger \cdot L_j
  - Y_d^{ij}\; \overline{d_{R}}_{i} \, \phi_2^\dagger \cdot Q_j
  - Y_u^{ij}\; \overline{u_{R}}_{i} \, Q_j^T \cdot \varepsilon \cdot \phi_2 + \mathrm{H.c.}~.
\end{equation}
To obtain the other three types, one or two of the occurences of $\phi_2$ in the above Lagrangian have to be exchanged for $\phi_1$, according to Table \ref{tbl::2hdm}.

\begin{table}
  \centering
  \begin{tabular}{cccc} \toprule
    Type & $u_R$ & $d_R$ & $e_R$ \\\midrule
    I & $\phi_2$ & $\phi_2$ & $\phi_2$ \\
    II & $\phi_2$ & $\phi_1$ & $\phi_1$ \\
    X & $\phi_2$ & $\phi_2$ & $\phi_1$ \\
    Y & $\phi_2$ & $\phi_1$ & $\phi_2$ \\
    \bottomrule
  \end{tabular}
  \caption{\label{tbl::2hdm} Couplings of the scalar doublets to right-handed fermions in the four $\mathbb{Z}_2$-symmetric 2HDMs.}
\end{table}

For the quartic scalar couplings $\lambda_i$ we have
\begin{equation} \label{eq:thdm_la}
  \mathcal{L}_{\mathrm{self}} = -\frac{\lambda_1}{2} (\phi_1^\dagger\phi_1)^2 -\frac{\lambda_2}{2} (\phi_2^\dagger\phi_2)^2 - \lambda_3(\phi_1^\dagger\phi_1)(\phi_2^\dagger\phi_2) - \lambda_4(\phi_1^\dagger\phi_2)(\phi_2^\dagger\phi_1)
                                -\left[\frac{\lambda_5}{2} (\phi_1^\dagger\phi_2)^2 + \mathrm{h.c.}\right]~.
\end{equation}
\texttt{La} is implemented in \texttt{base} as a vector of \texttt{std::complex<double>} with 5 entries.

\paragraph{mssm.h}
implements the class \texttt{mssm}, which is derived as \texttt{public} from \texttt{base}. A \texttt{mssm} object can be constructed by
\begin{lstlisting}
  mssm() : base<3,0>() {};
 
  mssm(const gauge<3> g_in, const yukawa Yu_in, const yukawa Yd_in,
       const yukawa Ye_in) : base<3,0>(g_in, Yu_in, Yd_in, Ye_in) {};
 
  mssm(const base<3,0> &X) : base<3,0>(X) {};
\end{lstlisting}
Note, that in the case of the MSSM, all quartic self-couplings of the scalars are determined by other couplings, namely gauge and Yukawa couplings.
Thus all entries of \texttt{La} are set to zero at construction.

The Yukawa couplings are fixed by the superpotential
\begin{equation} \label{eq:mssm_yuk}
  \mathcal{W_\mathrm{Yukawa}}^{\mathrm{MSSM}} = Y_e^{ij} \; {e_R}_i\, H_1\cdot \varepsilon \cdot L_j \quad + Y_d^{ij} \; {d_R}_i \, H_1\cdot \varepsilon \cdot Q_j + Y_u^{ij} \; {u_R}_i \, \cdot Q_j\cdot \varepsilon \cdot H_2 \;.
\end{equation} 

\paragraph{nusm.h}
implements the class \texttt{nusm} which corresponds to the SM extended by three generations of right-handed neutrinos.
\texttt{nusm} is derived as \texttt{public} from \texttt{nubase} and therefore has all members and member functions of \texttt{nubase}. A \texttt{nusm} object can be constructed by
\begin{lstlisting}
  nusm() : nubase<3,1>() {};
 
  nusm(const gauge<3> g_in, const std::complex<double> La_in,
       const yukawa Yu_in, const yukawa Yd_in, const yukawa Ye_in,
       const yukawa Yn_in, const yukawa Ka_in, const yukawa Mn_in)
     : nubase<3,1>(g_in, Yu_in, Yd_in, Ye_in, Yn_in, Ka_in, Mn_in)
     {La[0] = La_in;};
  
  nusm(const nubase<3,1> &X) : nubase<3,1>(X) {};
\end{lstlisting}
The conventions for the Yukawa couplings and self-couplings are those of eqs.~\eqref{eq:sm_yuk} and \eqref{eq:sm_la}.

The conventions for \texttt{Yn, Ka, Mn} are defined through
\begin{equation}\label{eq:sm_nu}
  \mathcal{L}_{\kappa}^{SM} = - Y_\nu^{ij}\; \overline{N}_i\, L_j^T \cdot \varepsilon \cdot \phi -\frac{1}{2}\, M^{ij}\; \overline{N}_i N^c_j  + \frac{\kappa^{ij}}{4} \bigl(\overline{L^c}_i\cdot \varepsilon \cdot \phi \bigr) \bigl(L_j^T\cdot \varepsilon \cdot \phi \bigr) + \mathrm{H.c.} \;.
\end{equation}

\paragraph{numssm.h}
implements the class \texttt{numssm} which is derived as \texttt{public} from \texttt{nubase} and therefore has all members and member functions of \texttt{nubase}. A \texttt{numssm} object can be constructed by
\begin{lstlisting}
  numssm() : nubase<3,0>() {};
  
  numssm(const gauge<3> g_in, const yukawa Yu_in, const yukawa Yd_in,
         const yukawa Ye_in, const yukawa Yn_in, const yukawa Ka_in,
         const yukawa Mn_in)
       : nubase<3,0>(g_in, Yu_in, Yd_in, Ye_in, Yn_in, Ka_in, Mn_in) {};
  
  numssm(const nubase<3,0> &X) : nubase<3,0>(X) {};
\end{lstlisting}
The conventions for the Yukawa couplings are the same as in eq.~\eqref{eq:mssm_yuk}.

For \texttt{Yn, Ka, Mn} we have
\begin{equation} \label{eq:mssm_nu}
    \mathcal{W}_{N} = \frac{1}{2}\, M^{ij}\; N_i\, N_j + Y_\nu^{ij}\; N_i\, L_j \cdot \varepsilon^T \cdot H_2 -\frac{\kappa^{ij}}{4} \bigl(L_i\cdot \varepsilon \cdot H_2\bigr) \bigl(L_j\cdot \varepsilon \cdot H_2\bigr)\;.
\end{equation}
As in the MSSM, all entries of \texttt{La} are set to zero at construction.

\subsection{\label{sec::auxiliary} Auxiliary classes}
In the header files \texttt{ckm.h} and \texttt{pmns.h} classes to obtain fermion masses and mixing angles from Yukawa matrices are implemented. The routines are \texttt{C++} implementations of the \texttt{Mathematica} package \texttt{MPT} that comes alongside \texttt{REAP}. Both classes therefore adopt the conventions of \texttt{MPT}. If you use one of these classes, please do not forget to cite \cite{Antusch:2005gp}.
\paragraph{ckm.h}
implements the class \texttt{ckm}.
A \texttt{ckm} object can be constructed by
\begin{lstlisting}
  ckm(const yukawa Yuin, const yukawa Ydin); // sets Yu and Yd
  ckm(const yukawa Yuin, const yukawa Ydin, const double vevin);
      // sets Yu, Yd and the vev
\end{lstlisting}
The routines that extract the quark observables from the Yukawa matrices are called by
\begin{lstlisting}
  void calculate();
\end{lstlisting}
Access to the results is given by the member functions
\begin{lstlisting}
  yukawa get_CKM(); // returns the CKM matrix
  Vector3d get_upyukawas(); // returns the central values y_u, y_c, y_t
  Vector3d get_downyukawas(); // returns the central values y_d, y_s, y_b
  Vector4d get_CKMparameters(); // return theta_12, theta_13, theta_23, delta_CP
  Vector3d get_upmasses(); // returns m_u, m_c, m_t
  Vector3d get_downmasses(); // returns m_d, m_s, m_b
  Vector3d get_upmasses(const double tanb); // return m_u, m_c, m_t using tan(beta)
  Vector3d get_downmasses(const double tanb); // returns m_d, m_s, m_b using tan(beta)
\end{lstlisting}
The convention for the CKM matrix is chosen such that
\begin{equation}
  Y_d = \operatorname{diag}(y_d,y_s,y_b)\,\cdot V_{\mathrm{CKM}}^\dagger\;,
\end{equation}
in the flavour basis where $Y_u$ is diagonal. The CKM parameters are defined as in the PDG \cite{Patrignani:2016xqp}.
The functions \verb|get_upmasses| and \verb|get_downmasses| are overloaded to also support models where the down-type quarks couple to another Higgs doublet than the up-type quarks,
such as the MSSM or the type-II 2HDM. To this end, the ratio between the two vacuum expectation values (vevs) of the two Higgs doublets
\begin{equation}
\tan\beta = \frac{v_2}{v_1}
\end{equation}
can be provided as an argument. Here $v_1$ is the vev of the doublet coupling to the down-type quarks and $v_2$ the vev of the doublet coupling to the up-type quarks.

\paragraph{pmns.h}
implements the class \texttt{pmns} which can be constructed by
\begin{lstlisting}
  pmns(const yukawa Min, const yukawa Yein); // sets M and Ye
  pmns(const yukawa Min, const yukawa Yein, const double vevin);
       // sets M, Ye and the vev
\end{lstlisting}
Note that $M$ is the mass matrix of the left-handed neutrinos.
The routines the extract the lepton observables are called by
\begin{lstlisting}
  void calculate();
\end{lstlisting}
Access to the results is given by the member functions
\begin{lstlisting}
  yukawa get_PMNS(); // returns the PMNS matrix
  Vector3d get_elyukawas(); // returns the central values y_e, y_mu, y_tau
  Vector3d get_numasses(); // returns the light neutrino masses
  Vector4d get_PMNSparameters(); // returns theta_12, theta_13, theta_23, delta_CP
  Vector3d get_elmasses(); // returns m_e, m_mu, m_tau
  Vector3d get_elmasses(const double tanb); // returns m_e, m_mu, m_tau using tan(beta)
  double get_betadecaymass(); // returns the effective beta decay mass for neutrinos
\end{lstlisting}
The convention for the PMNS matrix is chosen such that
\begin{equation}
  V_{\mathrm{PMNS}}^T \cdot m_\nu \cdot V_{\mathrm{PMNS}}
\end{equation}
diagonalises $m_\nu$ in the flavour basis where $Y_e$ is diagonal. The PMNS parameters are defined as in the 'standard convention' \cite{Patrignani:2016xqp}.

\paragraph{rundown.cpp}
The template function\texttt{rundown} provides routines that simplify the numerical integration of the RGEs. As it uses the \texttt{auto} functionality, \texttt{rundown} will not work with standards prior to \texttt{C++11}. The numerical integration is called with
\begin{lstlisting}
  rundown <Stepper, Model, Observer> (Model & state, double start,
           double end, double error = 1e-10, double firststep = 0.01);
\end{lstlisting}
\texttt{Stepper} needs to be of \texttt{odeint} Error Stepper type, e.g. \texttt{runge\_kutta\_fehlberg78}. See the examples in Section~\ref{sec::examples} or the \texttt{odeint} documentation for greater detail. \texttt{Model} refers to an \texttt{RGE++} model (\texttt{sm}, \texttt{mssm}, \texttt{tdhmi/ii/x/y}, \texttt{nusm}, \texttt{numssm}) and \texttt{state} to an instantation thereof. \texttt{Observer} is a struct or class that provides an operator of the form \texttt{void operator()(const \& Model, double t)} that is called at each step of the numerical integration providing access to intermediate results. A possible example is given in \texttt{numssm\_example.cpp}. \texttt{Observer} can also be \texttt{void}, if intermediate results are not necessary.

The arguments \texttt{start} and \texttt{end} denote the starting point $\ln\mu_{\mathrm{start}}$ and end point $\ln\mu_{\mathrm{end}}$ of the numerical integration. If \texttt{start} $>$ \texttt{end}, i.e. when going from higher to lower renormalisation scales, \texttt{rundown} will determine the thresholds at which particles need be integrated out using the \texttt{logthresholds()} functionality of \texttt{Model}. At these thresholds \texttt{rundown} will integrate out particles by using the \texttt{integrate\_out(int)} member function of \texttt{Model}. If \texttt{end} $<$ \texttt{start}, i.e. when going towards higher renormalisation scales, no particles are integrated out.\\
The arguments \texttt{error} and \texttt{firststep} control the numerical error at each step of the integration and the size of the first integration step. If no value is given they default to $10^{-10}$ and $0.01$, respectively. From the second integration step onwards, the step size is adjusted by the algorithms of \texttt{odeint} based on the value of \texttt{error}.

\section{\label{sec::examples}Examples and usage}
In the following, we discuss the basic usage of models implemented in \texttt{RGE++} to evolve parameters from one scale to another.

\subsection{\label{sec::sm_example} A SM example}
In \texttt{sm\_example.cpp} the parameters of the SM are given at the Z-Boson mass and we employ the beta functions to extrapolate them to higher energy scales. 
To build the example call \texttt{make sm\_example} from the main folder. Then, the program can be called from the command line via
\begin{lstlisting}
  ./examples/sm_example
\end{lstlisting}
Since we have not given any additional options to the program the output should look like
\begin{lstlisting}[keywordstyle=\color{blue}]
  SM parameters at 3000 GeV and 2 loops:
  gauge couplings: 0.47077 0.63383  1.0008
  up-type yukawas: 6.3691e-06   0.002976    0.83158
  down-type yukawas: 1.3614e-05 0.00025413   0.013092
  charged lepton yukawas: 2.8666e-06 0.00060516   0.010287
  ckm parameters:   0.22704 0.0038204  0.043031     1.143
  Higgs quartic coupling: (0.50749,-8.5389e-59)
\end{lstlisting}
This example program also accepts other choices of the energy scale or the loop orders which should be used for running the parameters.
The SM parameters at, e.g., 5000 GeV using three-loop accuracy can be obtained by calling
\begin{lstlisting}
  ./examples/sm_example 5000 3
\end{lstlisting}

Having established the general idea of the example, we are now in the position to discuss the corresponding source code.
The first two lines of the \texttt{main} function parse the options given to the program via the command line. The first argument denotes the final renormalisation scale. If no argument is given it defaults to 3000 GeV. The
second argument stands for the number of loops included in the beta functions. It defaults to 2 but can be chosen to be as high as 4, as the gauge coupling beta functions are implemented at this order,
\begin{lstlisting}
  double scale = (argc == 2) ? atoi(argv[1]) : 3000.;
  int nloops = (argc == 3) ? atoi(argv[2]) : 2;
\end{lstlisting}
Then the initial renormalisation scale is defined at which the input parameters are given
\begin{lstlisting}
  double MZ(91.1876);
\end{lstlisting}
The model parameters are \texttt{eigen3} matrix types. The down-type Yukawa matrix $Y_d$ e.g. is initialized and defined as \footnote{Please cite \cite{Deppisch:2018flu} and references therein if you use these numbers.}
\begin{lstlisting}
  yukawa Yd; 
  Yd << 0.0000166293, 0,0,0, 0.000310436, 0,0,0, 0.0164568;
\end{lstlisting}
The non-zero numbers are the down-, strange- and bottom-quark Yukawa couplings in the basis where the down-type Yukawa matrix is diagonal.
For details on how to define and manipulate  \texttt{eigen3} matrix types see the \texttt{eigen3} manual. In a flavour basis where $Y_u$ is diagonal, $Y_d$ takes the form
\begin{equation}
    Y_d = \operatorname{diag}(y_d,y_s,y_b) \cdot V_{\mathrm{CKM}}^\dagger\;.
\end{equation}
This rotation is performed by the line
\begin{lstlisting}
  Yd = Yd*ckm(th12,th13,th23,phi).adjoint();
\end{lstlisting}
Note that the function \texttt{ckm} returns the CKM matrix in the conventions of the PDG \cite{Patrignani:2016xqp} taking the mixing angles and the CP phase as input and is defined at the beginning of \texttt{sm\_example}.

Using these input parameters we can now construct the \texttt{sm} class object \texttt{values} which contains both the RGEs and the input parameters
\begin{lstlisting}
  sm values(g, lambda, Yu, Yd, Ye, nloops);
\end{lstlisting}
For the numerical integration an \texttt{ODEint} stepper needs to be defined \footnote{See the \texttt{ODEint} documentation for other steppers. In our tests all adaptive steppers performed equally well.}
\begin{lstlisting}
  using namespace boost::numeric::odeint;
  typedef runge_kutta_fehlberg78< sm, double, sm, double, vector_space_algebra > stepper;
\end{lstlisting}
The integration is then performed by
\begin{lstlisting}
  int steps = integrate_adaptive(make_controlled<stepper>(1E-10, 1E-10),
                                 sm(), values, log(MZ), log(scale), 0.01);
\end{lstlisting}
The \texttt{sm} class object \texttt{values} now contains the SM parameters at the renormalisation scale \texttt{scale}. The argument \texttt{make\_controlled<stepper>( 1E-10 , 1E-10 )} concerns the error control
and the last argument \texttt{0.01} defines the size of the first step of the integration. For calculating RGEs, these are sensible values from our experience.
Note that the beginning (\texttt{log(MZ)}) and the end (\texttt{log(scale)}) of the integration need to be given as logarithms since the beta functions are implemented as the logarithmic derivatives of the
renormalised theory parameters with respect to the renormalisation scale.

To extract the physical observables, i.e. fermion masses and their mixing angles,  from the Yukawa matrices we can use the \texttt{ckm} and \texttt{pmns} classes. A \texttt{ckm} object is defined by
\begin{lstlisting}
  class ckm quarks(values.Yu, values.Yd);
\end{lstlisting}
and the quark masses and CKM parameters are calculated by calling
\begin{lstlisting}
  quarks.calculate();
\end{lstlisting}
They can be accessed as shown in the final lines of the example
\begin{lstlisting}
  std::cout << "up-type yukawas: "
            << quarks.get_upyukawas().transpose() << std::endl;
\end{lstlisting}

\subsection{\label{sec::sm_plot_example} SM example suitable for plotting}
A further SM example is given in \texttt{running\_plot.cpp}. It can be built by invoking \texttt{make running\_plot} from the main folder.
The setup is similar as in the previous example, however, instead of an adaptive stepsize, it uses a constant one.
In this case, the integration is performed by
\begin{lstlisting}
  int steps = integrate_const(stepper(), sm(), values, log(MZ),
                              log(MGUT), 1., write_out());
\end{lstlisting}
where the last argument of \texttt{integrate\_const} is a structure with an overloaded operator \texttt{()} that gets called at each step.
It is implemented as
\begin{lstlisting}
  struct write_out {
    void operator()(const sm &x, double t) {

      // calculate quark yukawas
      ckm quarks(x.Yu, x.Yd);
      quarks.calculate();

      // calculate lepton yukawas
      yukawa zero;
      pmns leptons(zero, x.Ye);
      leptons.calculate();
    
      // write to standard output
      std::cout << exp(t) << "   "
	      << alpha(x.g[0]) << "   "
      	      << alpha(x.g[1]) << "   "
      	      << alpha(x.g[2]) << "   "
	      << alpha(quarks.get_upyukawas()[2]) << "   "
	      << alpha(quarks.get_downyukawas()[2]) << "   "
	      << alpha(leptons.get_elyukawas()[2]) << std::endl;
    }
  };
\end{lstlisting}
and prints the $\alpha_i = g_i^2/\left(4\pi\right)$ to standard output.

\begin{figure}
\centering
\includegraphics{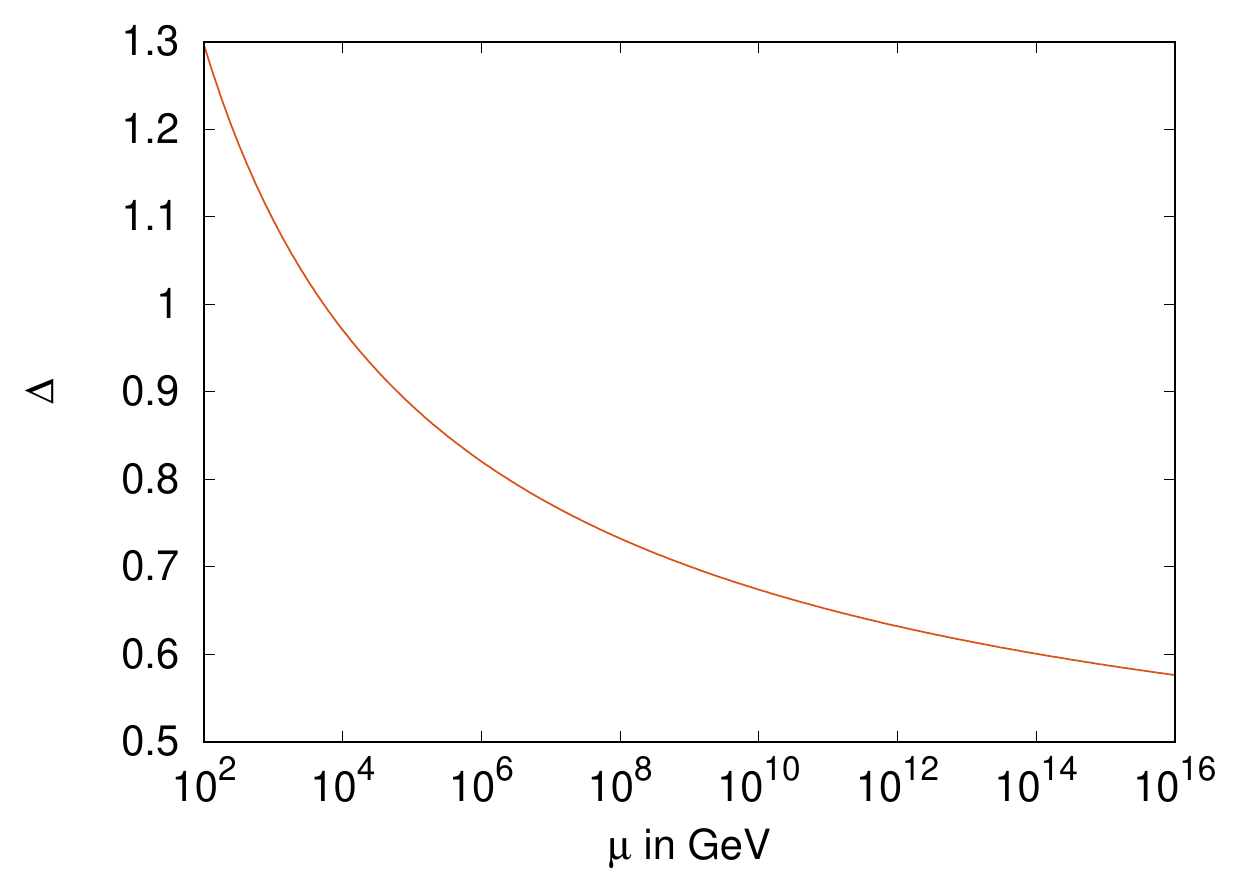}
\caption{\label{plt::delta} $\Delta$ plotted over the renormalisation scale. For energies above $10^4$~TeV, the three-loop terms have a bigger impact than the four-loop ones. Still,
the four-loop contributions are at least half the size of the three-loop ones for the renormalisation scales under consideration.}
\end{figure}

In \cite{Davies:2019onf} it was observed, that the four-loop contributions to the
beta function of the strong coupling, $\beta_3$, is larger than the three-loop contributions by a factor of 1.27 for $\mu = M_Z$. This seemingly bad perturbative convergence
is due to an accidental cancellation in the three-loop terms \cite{Davies:2019onf}, which is not present in the four-loop terms. Using the code of this example, this can be seen from Fig.~\ref{plt::delta}, where the dependence of
\begin{align}
\Delta = \frac{|\alpha_3^{(4l)} - \alpha_3^{(3l)}|}{|\alpha_3^{(3l)} - \alpha_3^{(2l)}|}
\end{align}
on the renormalisation scale is shown.
Indeed, for renormalisation scales below $10^4$~GeV, the four-loop contributions have a larger impact than the three-loop ones. However, for larger scales,
the accidental cancellation in the three-loop terms is lifted and the perturbative convergence improves.

\subsection{\label{sec::mssm_example} A SUSY GUT example}
Another useful application is to evolve the RGEs in the MSSM including right-handed neutrinos from a high scale down to $M_Z$, where it can be matched onto the SM. Such an example is given in the file \texttt{numssm\_example.cpp}.
To build the example call \texttt{make numssm\_example} from the main folder.
Let us assume the Yukawa and gauge couplings are set by some (unrealistically oversimplified) GUT scenario, i.e.
\begin{equation}
g_1 = g_2 = g_3 = 0.7, \qquad Y_u = Y_\nu = Y_1 + Y_2 \qquad Y_d = Y_e = Y_1 - Y_2, 
\label{eq::GUT}
\end{equation}
with some numerical Yukawa matrices $Y_1$ and $Y_2$ and the right-handed neutrino mass matrix $M = 10^{14}\cdot Y_1$, which gives rise to neutrino masses but
does not influence the running of the gauge couplings and Yukawa matrices. These conditions hold at $M_{GUT} = 2\times 10^{16}\ \mathrm{GeV}$.

The code \texttt{numssm\_example.cpp} starts with defining the initial and final renormalisation scales
\begin{lstlisting}
  double MGUT = 2.e16;
  double MZ = 91.1876;
\end{lstlisting}
The Yukawa matrices and the gauge couplings are defined by
\begin{lstlisting}
  yukawa zero, Y1, Y2;
  gauge g;

  Y1 << 1.e-4, 0., 0.,  0., 2.e-2, 0.,  0., 0., 0.4;
  Y2 << 0., -2.e-3, 3.e-2,  -2.e-3, 0., 0.5,  3.e-2, 0., 0.5;
  g << 0.7, 0.7, 0.7;
\end{lstlisting}
Using those input parameters, and following Eq.~(\ref{eq::GUT}) we construct a \texttt{numssm} object via
\begin{lstlisting}
  numssm values(g, Y1 + Y2, Y1 - Y2, Y1 - Y2, Y1 + Y2, zero, Y1*1.e14); 
\end{lstlisting}
The matrix \texttt{zero} contains zeros in each entry and is used to set the Wilson coefficient of the Weinberg operator $\kappa$ to zero at the beginning.
 
Again, the \texttt{ODEint} stepper functions have to be set up for the numerical integration
\begin{lstlisting}
  using namespace boost::numeric::odeint;
  typedef runge_kutta_fehlberg78< numssm, double, numssm, double, vector_space_algebra > stepper;
\end{lstlisting}
This time, however, we will use the \texttt{rundown} template which not only performs the integration but also determines the see-saw scale (i.e. the mass scales of the right-handed neutrinos) and integrates out the right-handed neutrinos at tree-level, when coming across these see-saw scales.
\begin{lstlisting}
  rundown<stepper,numssm,void> foo(values, log(MGUT), log(MZ));
\end{lstlisting}
Note, that although the name suggests otherwise, \texttt{rundown} works `both ways', i.e. both from the high to the low as well as from the low to high renormalisation scale. Only when evolving to a lower renormalisation scale, however, it will check for thresholds and integrate out the relevant particles using the \texttt{integrate\_out()} routine of the respective model (\texttt{numssm} in this case).
  
Since the MSSM contains two Higgs doublets, also the ratio of their respective vevs, $\tan\beta = v_2 / v_1$, has to be defined in order to calculate the fermion masses.
\begin{lstlisting}
  double tanb(60.);
\end{lstlisting}
As before the quark masses and their mixing can be extracted from the Yukawa matrices by
\begin{lstlisting}
  class ckm quarks(values.Yu, values.Yd);
  quarks.calculate();
\end{lstlisting}
In the lepton sector we additionally need the mass matrix of the left-handed neutrinos which is calculated from the Wilson coefficient of the Weinberg operator $\kappa$.
\begin{lstlisting}
  yukawa ML = values.getML(tanb);   // left-handed neutrino mass matrix
  class pmns leptons(ML, values.Ye);
  leptons.calculate();
\end{lstlisting}
The results can be written to the standard output similar to
\begin{lstlisting}
  std::cout << "up-type masses (GeV): " << quarks.get_upmasses(tanb).transpose() << std::endl;
\end{lstlisting}    
The complete output should then look like
\begin{lstlisting}[keywordstyle=\color{blue}]
  nuMSSM parameters at MZ:
  gauge couplings: 0.45494 0.63045  1.1092
  ckm parameters:  0.95698 0.058163 0.030761   3.1416
  up-type masses (GeV): 0.6205 7.3594 170.36
  down-type masses (GeV): 0.0053232   0.20148    2.2537
  PMNS parameters:    1.5057  0.037873 0.0012002    3.1416
  neutrino masses  (eV): 0.00037949    0.01031     3.4087
  charged lepton massess  (GeV): 0.0025152  0.095236    1.3424
\end{lstlisting}    

\subsection{\label{sec::adding} Adding additional Models}
In this example we outline the necessary steps for implementing additional models. As an example, we implement the RGEs for the SM with an additional singlet scalar field $s$ such that the scalar sector of the theory takes the form
\begin{equation} \label{eq:nsm_la}
  \mathcal{L}_{\mathrm{self}} = -\frac{\lambda_1}{4} (\phi^\dagger\phi)^2  -\frac{\lambda_2}{2} (\phi^\dagger\phi)\ (s^\dagger s)   -\frac{\lambda_3}{4} (s^\dagger s)^2 ,
\end{equation}
where additional couplings are forbidden by an ad-hoc $\mathbb{Z}_2$ symmetry. For reference of such models see e.g. \cite{Schabinger:2005ei,Patt:2006fw,Bowen:2007ia}.

Regarding the fermion sector, this model is identical to the SM, since the singlet couples neither to quarks nor leptons. Therefore, the RGEs for the Yukawa couplings are the same as in the SM at leading order.
The same holds also for the gauge couplings. We only need to adapt the RGEs in the scalar sector, which can be obtained from the general results in \cite{Machacek:1984zw,Poole:2019kcm}.

Let us start this example with the header file \texttt{nsm.h} in the \texttt{models} folder. The class definition of the \texttt{nsm} class is
\begin{lstlisting}
  class nsm : public base<3,3> {
  
   public:
    nsm() : base<3,3>() {};

    nsm(const gauge<3> g_in, const std::complex<double> La1_in,
      const std::complex<double> La2_in, const std::complex<double> La3_in,
      const yukawa Yu_in, const yukawa Yd_in, const yukawa Ye_in)
      : base<3,3>(g_in, Yu_in, Yd_in, Ye_in, 1)
      { La << La1_in, La2_in, La3_in; };

    nsm(const base<3,3> &X) : base<3,3>(X) {};
 
    // contains the RGEs
    void operator()(const nsm &X, nsm &dX, const double);
  };
\end{lstlisting}
The first line lets the \texttt{nsm} class inherit the properties of the \texttt{base} class specifying the number of gauge couplings (3) and the number of scalar couplings (also 3: $\lambda_1$, $\lambda_2$, $\lambda_3$).
After the \texttt{public} statement three constructors are defined. The first one, setting all couplings to 0, is inherited from \texttt{base}.
The second one allows one to construct an \texttt{nsm} object using three complex variables as input for $\lambda_1$, $\lambda_2$, $\lambda_3$ in addition to the Yukawa couplings,
which are then stored in the vector of self-couplings \texttt{La}.
In this example, we will only consider one-loop RGEs, thus the variable (\texttt{nloops}) is set to one.
For the function \texttt{void operator()} only the type of the input parameters has been changed in comparison to \texttt{sm.h}.

After the class definition there is an additional piece of code copied from \texttt{sm.h} that defines a struct that returns the modulus of the largest coupling in the model. This code does not need to be changed as long as there are no additional Yukawa couplings. In that case it should be straightforward to include theses additional couplings in the calculation of the largest coupling.

Now we are ready to set up the actual RGEs in the file \texttt{nsm.cpp}. Again, for the gauge and Yukawa couplings we can simply copy the SM ones from from \texttt{sm.cpp}. The code that needs some further adjustments is marked with a comment reading \texttt{NEW CODE}. The RGEs for the scalar couplings read
\begin{lstlisting}
  self<3> La2 = X.La.cwiseAbs2();

  dX.La[0] = loopfactor*((12.)*X.La[0]*Yd2Tr + (-24.)*Yd4Tr
             + (4.)*X.La[0]*Ye2Tr + (-8.)*Ye4Tr + (12.)*X.La[0]*Yu2Tr
             + (-24.)*Yu4Tr + (-9./5.)*X.La[0]*g2[0]
             + (-9.)*X.La[0]*g2[1] + (9./5.)*g2[0]*g2[1]
             + (27./50.)*g4[0] + (9./2.)*g4[1]
             + 6.*La2[0] + 6.*La2[1]);
  dX.La[1] = loopfactor*( 4.*X.La[0]*X.La[1] + 4.*X.La[1]*X.La[2] );
  dX.La[2] = loopfactor*( 6.*La2[2] + 6.*La2[1]);
\end{lstlisting}
The first line defines a convenient abbreviation for the modulus squared of the vector self-couplings \texttt{La}. The second assignment resembles the RGE for $\lambda_1$. The part that stems from the interactions with the SM particles is, of course, the same as in \texttt{sm.cpp}. But there is an additional term from diagrams where singlet scalars appear in the loops. The next two assignments contain the RGEs of $\lambda_1$ and $\lambda_2$. We further deleted all higher loop orders that appear in \texttt{sm.cpp} for the sake of transparency.

The application of this model is demonstrated in the \texttt{nsm\_example.cpp} file which is an adapted version of \texttt{sm\_example.cpp}. Upon compilation and running the command line prompt should look like
\begin{lstlisting}[keywordstyle=\color{blue}]
  nSM parameters at  1000 GeV:
  Higgs self-couplings: (0.65818637,0)  (1.0515135,0) (0.19774487,0)
\end{lstlisting}
Note that the numbers are chosen arbitrarily for demonstration purposes.

\section{\label{sec::sum}Summary and Outlook}
We present the \texttt{C++} library \texttt{RGE++} for solving RGEs in theories with multiple couplings numerically. This concerns in particular the Standard Model up to four loops, as well as
two-Higgs-doublet models and the MSSM up to three loops. Furthermore, \texttt{RGE++} supports models with right-handed neutrinos and provides routines for decoupling them from the evolution.
We provide three examples showcasing the usage and discuss the implemented classes in detail. The source code and further examples is available from \url{https://github.com/Herren/RGEpp}.
The template-based structure gives the user the flexibility to implement higher-order corrections or new models easily.

\section*{Acknowledgements}
We thank Joshua Davies, Marvin Gerlach and Matthias Linster for testing the code, as well as Josha Davies, Matthias Steinhauser and Anders Eller Thomsen for careful reading of the manuscript.
TD would like to thank Martin Spinrath and Stefan Schacht for collaboration on the project initiating the development of \texttt{RGE++}.
This research was supported by the Deutsche Forschungsgemeinschaft (DFG, German Research Foundation) under grant 396021762 - TRR 257 "Particle Physics Phenomenology after the Higgs Discovery".
FH acknowledges the support by the Doctoral School "Karlsruhe School of Elementary and Astroparticle Physics: Science and Technology".

\printbibliography
\end{document}